# Start-phase control of distributed systems written in Erlang/OTP

Péter Burcsi
Eötvös Loránd University
Faculty of Informatics
Department of Computer Algebra
email:
peter.burcsi@compalg.inf.elte.hu

Attila Kovács
Eötvös Loránd University
Faculty of Informatics
Department of Computer Algebra
email:
attila.kovacs@compalg.inf.elte.hu

Antal Tátrai
Eötvös Loránd University
Faculty of Informatics
Department of Computer Algebra
email:
antal.tatrai@compalg.inf.elte.hu

**Abstract.** This paper presents a realization for the *reliable* and *fast* startup of distributed systems written in Erlang. The traditional startup provided by the Erlang/OTP library is sequential, parallelization usually requires unsafe and ad-hoc solutions. The proposed method calls only for slight modifications in the Erlang/OTP `stdlib` by applying a system dependency graph. It makes the startup safe, quick, and it is equally easy to use in newly developed and legacy systems.

## 1 Introduction

A distributed system is usually a collection of processors that may not share memory or a clock. Each processor has its own local memory. The processors

---

in the system are connected through a communication network. Communication takes place via messages [11]. Forms of messages include function invocation, signals, and data packets. Computation based models on message passing include the actor model and process algebras [4]. Several aspects of concurrent systems written in message passing languages have been studied including garbage collection [2], heap architectures [7], or memory management [8]. Startup concurrency is an area not fully covered yet.

Why is the investigation of the startup phase important?

- During system and performance testing, when the system is frequently started and stopped, fast startup might be beneficial.
- Critical distributed systems often have the maintainability requirement of $99.999$ availability, also known as the "five nines". In order to comply with the "five nines" requirement over the course of a year, the total boot time could not take more than $5.25$ minutes. In practice, due to the maintenance process, every system has a planned down time. In this case a fast and reliable startup is a must.
- The startup time is not the only reason to study the startup phase. In most product lines the requirements (and therefore the code) alter continuously. The changes may influence the code structure, which may affect the execution order of the parts. Although the code can often be reloaded without stopping the system, the changes may influence the startup. The challenge is to give a generic solution which supports reliable, robust and fast startup even when some software and/or hardware parts of the system had been changed.

In this paper we focus on the distributed programming language Erlang. Erlang was designed by the telecommunication company Ericsson to support fault-tolerant systems running in soft real-time mode. Programs in Erlang consist of functions stored in modules. Functions can be executed concurrently in lightweight processes, and communicate with each other through asynchronous message passing. The creation and deletion of processes require little memory and computation time. Erlang is an open source development system having a distributed kernel [6].

The Erlang system has a set of libraries that provide building primitives for larger systems. They include routines for I/O, file management, and list handling. In practice Erlang is most often used together with the library called the Open Telecom Platform (OTP). OTP consists of a development system platform for building, and a control system platform for running telecommunication applications. It has a set of design principles (behaviours), which together with middleware applications yield building blocks for scalable robust real time systems. Supervision, restart, and configuration mechanisms are provided. Various mechanisms, like an ORB, facilitate the development of CORBA based management systems. Interfaces towards other languages include a Java interface, an interface allowing Erlang programs to call C modules, etc. These interfaces are complemented with the possibility of defining IDL interfaces, through which code can be generated. The number of Erlang/OTP applications and libraries is continuously increasing. There are for example SNMP agents, a fault tolerant HTTP server, a distributed relational database called Mnesia, etc. One of the largest industrial applications developed in Erlang/OTP is the AXD 301 carrier-class multi-service (ATM, IP, Frame-relay, etc.) switching system of Ericsson. It is a robust and flexible system that can be used in several places of networks. It has been developed for more than 10 years (for an early announcement of the product, see [5]), resulting in a long product line. It contains several thousand Erlang modules and more than a million lines of code.

How is the startup of an Erlang application performed? The traditional startup provided by the Erlang/OTP library is sequential. It was not designed to start as quickly as possible, no special attention was paid to the possibility of parallelizing the different operations performed during startup. The only order imposed is due to the explicit dependencies described in the application configuration files. Technically, the reason of the sequential startup is that each process performing an OTP behaviour sends an ACK (acknowledge) signal to its parent only after the whole initialization process is finished. It means that each process has implicit preconditions. In the concurrent case, maintaining these preconditions is a fundamental problem. The proposed solution enables the concurrent startup and provides an Erlang/OTP extension for describing and realizing preconditions between behaviour processes. Hence the startup will not only be fast but remains reliable as well. The use of conditions to construct dependency graphs to manage the order of startup bears a resemblance to the mechanism used by Apple's MacOSX StartupItems. Each StartupItem includes a properties list of items that provides/requires/uses other items, which are used by the SystemStarter to build a soft dependency graph controlling the order of starting items [10]. There does not exist any such mechanism in Erlang/OTP.

Of course, the startup times do not only depend upon the dependencies among the applications and the degree to which these startup activities can be parallelized. The startup times are affected by several other factors, probably the most significant being disk I/O times and latencies, the time spent unnecessarily searching for hardware elements, disks, appropriate files to load, etc. In a particular system measurements are needed to find where the time goes on for startup. In this paper we do not focus on a particular system, we give instead a general solution for performing fast and reliable startup in any Erlang/OTP systems. It means that dependencies among the applications must be given in advance. These can be determined by the system designers.

The paper is structured as follows. For completeness, Section 2 contains the basic description of Erlang/OTP features and concepts. In Section 3 the basic idea of the concurrent startup of Erlang applications is presented. Section 4 deals with the details of the proposed solution presenting a prototype. The measurements of the performance of our prototypes are written in Section 5 and finally the authors write a show conclusion in Section 6.

## 2 Erlang/OTP

In this section a short description of Erlang/OTP concepts is given. The overview begins with a few Erlang language features, then OTP design principles and the startup mechanism are discussed. For a full description of Erlang with many examples the authors refer to the books [1, 3] and to the on-line documentation [6].

### 2.1 Code structure and execution

The code written in Erlang is structured as follows:

- *Functions* are grouped together in source files called *modules*. Functions that are used by other modules are exported, modules that use them must import them. Or alternatively, have to use the `apply(Mod, Fun, Arg)` built-in function, or the `module_name:function_name (args)` form.
- Modules that together implement some specific functionality, form an *application*. Applications can be started and stopped separately, and can be reused as parts of other systems. Applications do not only provide program or process structure but usually a directory structure as well. There is a descriptor file for each application containing the module names, starting parameters and many other data belonging to the application.
- A *release*, which is the highest layer, may contain several applications. It is a complete system which contains a subset of Erlang/OTP applications and a set of user-specific applications. A release is described by a specific file, called release resource file. The release resource file can be used for generating `boot_scripts` for the system, and creating a *package* from it. After creating a `boot_script`, the system is able to start. First, the Erlang kernel is loaded. Then, a specific `gen_server` module (`application_controller`) is started. This module reads the application descriptor files sequentially, and creates a process called `application_master` for each application. The `application_master` starts the corresponding application, and sends an ACK signal back when the start is finished. Thus, as it was mentioned earlier, the Erlang/OTP startup is sequential.

The central concept of the execution is the process. As Erlang is message-oriented, executing Erlang code means creating strongly isolated processes that can only interact through message passing. Process creation, which is a lightweight operation, can be performed using the `spawn` family of functions. These functions create a parallel process and return immediately with the process ID (Pid). When a process is created in this way, we say that it is spawned. Erlang messages are sent in the form `Pid!Msg` and are received using `receive`.

## 2.2 Design principles

One of the most useful features in OTP is to have a pre-defined set of design patterns, called *behaviours*. These patterns were designed to provide an easy-to-use application interface for typical telecommunication applications such as client-server connections or finite state machines. In order to realize highly available and fault-tolerant systems, OTP offers a possibility to structure the processes into *supervision trees.*

### 2.2.1 Supervision trees

A principal OTP concept is to organize program execution into trees of processes, called supervision trees. Supervision trees have nodes that are either *workers* (leaves of the tree) or *supervisors* (internal nodes). The workers are Erlang processes which perform the functionality of the system, while supervisors start, stop, and monitor their child processes. Supervisor nodes can make decisions on what to do if an error occurs. Supervision tasks have a generic and a specific part. The generic part is responsible e.g. for the contact with the children, while the specific part defines (among other things) the restarting strategy. It is desirable that workers have a uniform interface, therefore OTP defines several behaviours with the same communication interface.

### 2.2.2 Behaviours

Behaviours, like every design pattern, provide a repeatable solution to commonly occurring problems. For example, a large number of simple server applications share common parts. Behaviours implement these common parts. A server code is then divided into a generic and a specific part. The generic part might contain the main loop of the server that is waiting for messages, and the specific part of the code contains what the server should do if a particular message arrives. In practice, only a *callback module* has to be implemented. OTP expects the existence of some functions (e.g. `handle_call`) in this module. As an example, several callback functions can be implemented for the complete functionality of an application, but the most important ones are: `start/2`, `stop/1`.

Let us summarize the most significant OTP behaviours: `gen_server`, `gen_fsm`, `gen_event` and `supervisor`. Each of them implements a basic pattern. The `gen_server` is the generic part of a server process, the `gen_event` is the generic part of event handling, the `gen_fsm` is the generic part of finite state machines. The `supervisor` behaviour is the generic part of the supervisor nodes of the supervision tree. Its callback module only contains the function `init(Arg)`, in which the children and the working strategy of the node can be specified. The `gen_server` behaviour also defines higher level functions for messaging, such as the synchronous (*call*) or asynchronous (*cast*) messages.

## 3 The basic idea of the solution

Let us suppose that we have an Erlang system and we plan to make the startup concurrent. If we use the `spawn` function instead of the built-in methods of supervisor child-starting then the spawned processes run parallelly, but the Erlang/OTP supervisor monitoring mechanism – one of the strongest Erlang/OTP features – is lost. Omitting the ACK mechanism from the built-in child-starting process would mean a deep redesign and reimplementation of the OTP (the ACK mechanism corresponds to sequential child-starting). Also, sequential start determines an order between processes, which would vanish using a too naive way of parallelization. Therefore an alternative parallel ordering is required to avoid dead-locks and startup crashes.

In the light of the previously mentioned properties we define our guidelines:

- The supervision tree structure, as well as other functionalities, must be preserved.
- The startup must be reliable and fast (faster than sequential).
- Only "small" modifications are permitted in the Erlang/OTP `stdlib`.

### 3.1 The dependency graph

In this subsection we consider dependence relations between modules and introduce the notion of dynamic dependency graphs.

In order to preserve the supervisor tree structure, we define *condition*s. Conditions represent the startup state of modules. A condition related to a module is false while the module's startup is being processed (or has yet to begin) and set to true when the corresponding startup has been finished. At the beginning of the startup all conditions are false. Conditions that the startup of another module depends on are called the preconditions of that module. A process can only start if all its preconditions are true. We can represent these relations in a dependency graph. Modules (or corresponding conditions) are the vertices, dependence between modules (or preconditions) are the directed edges in this graph.

When a behaviour module starts instruction defined by the first user (which is also the first that can imply preconditions) is the first instruction of the module's `init` function. Therefore the verification of the preconditions and setting up the completed conditions to true have to insert immediately before and after executing the `init` function.[1]

Dependency graphs are widely used in computer science. For example, dependency graphs are applied in the startup of the Mac OSX operating system [10] or in compiler optimization [9]. Moreover, a dependency graph is created when the Erlang boot script is generated from a given application.

However, there is a significant difference between the graphs above and our graph. The same Erlang software start up in different ways in different environments, therefore module parameters, execution and dependencies can vary and should be handled dynamically for full performance. In order to keep Erlang's robustness, we add one more guideline to the above:

- The dependency graph should be dynamic.

---

[1] We remark that during the sequential startup there exist implicit preconditions which are described in the hierarchy of the supervisor trees.

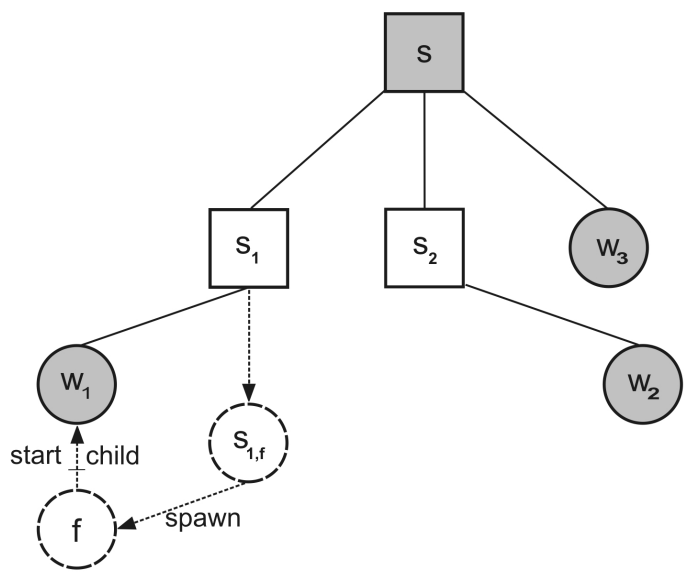

Figure 1: Inserting a *dummy* supervisor node for (1) preserving the supervisor's restarting behaviour, and (2) enabling fast parallel start-up. Supervisors are denoted by squares, permanent processes by continuous border, and temporary processes by dashed border. In the middle of the tree one can see the living processes ($s_2, w_2$) after termination of the dummy functions.

## 3.2 Concurrent startup of a supervisor's children

We also propose an Erlang trick that enables starting processes in a concurrent way. Here we can set which nodes should start concurrently. When a supervisor process s starts a child process $w_1$, the system starts a *dummy* (or *wrapper*) node $s_1$ instead. Then, the dummy process $s_1$ starts a simple function $s_{1,f}$ (which just calls a spawn function) and sends an ACK message back immediately to its parent s. Consequently, the next child $w_2$ of the supervisor node s can start. So far function $s_{1,f}$ has spawned function f. The spawned function f starts process $w_1$ and attaches it into the dummy process $s_1$ using the supervisor::start_child function[2]. The already started dummy process $s_1$ runs independently (parallel) from the other parts of the system. The ad-vantage of this method is that if process $w_1$ has a blocking precondition then only $w_1$ is waiting instead of the whole system. Figure 1 shows the described supervision hierarchy after start. The dummy supervisor node's restart strategy can be set in such a way that a crashing child results in the termination of the dummy supervisor. Thus the connection between $s$ and $w_1$ is preserved.

---

[2]The supervisor::start_child function takes effect just after the ACK message has been sent back.

The following code fragment shows the concurrent startup of a supervisor's children.

```
-module(dummy_sup_tree).

dummy_child({Tree_id, Child_spec}) ->
  spawn(dummy_sup_tree, child_starter, [{Tree_id, Child_spec}]),
  {ok, self()}.

child_starter({Tree_id, Child_spec}) ->
  supervisor:start_child(Tree_id, Child_spec),
  ok.

start_link({Child_spec}) ->
  supervisor:start_link(dummy_sup_tree, [{Child_spec}]).

init([{Child_spec}]) ->
  Sup_flags = {one_for_one, 0, 1},
  {ok,
   {Sup_flags,
    [
     {dummy_child_id, {dummy_sup_tree, dummy_child,
         [{self(), Child_spec}]}, temporary, brutal_kill,
         worker, [dummy_sup_tree, generic_server]}
    ]
   }
  }.
```

## 4  The solution's prototype

In this section we give the details of our solution by describing the skeleton of a prototype. We discuss the implementation of the dependency graph and how the Erlang boot script, the supervisors' `init` function, `stdlib` modules, etc. should be modified.

### 4.1 Realization of the dependency graph

The dependency graph is implemented as a module called `release_graph`. This module implements and exports the following functions: `get_conditions`, `get_preconditions` and `get_condition_groups`.

The `get_conditions` function returns a list of pairs. Each pair consists of a module name (with parameters) and a condition name.

   { {Mod, Args} , condition_name } .

We note that the function tag of the MFA (Module-Function-Arguments triplet) may be omitted, since it is always the `init` function of the module. The function `get_conditions` corresponds to the vertices of the dependency graph. Observe that a condition corresponds to a module together with parameters rather than a module, in accordance with our dynamic dependency graph guideline. In general, the `Args` parameter can be an actual parameter value or `undefined`. In the latter case the condition describes the module's startup with arbitrary parameters.

The `get_preconditions` function also gives a list. The elements of the list have the following structure:

   { { Mod, Args } , [ condition_names ] } .

The function corresponds to the edges of the dependency graph. When a module's `init` function is called then the validity of the conditions in the list must be tested. Once again, the `Args` parameter can be `undefined` meaning that the startup of this module with any parameters has to wait until all conditions in the list become true.

The third function facilitates the management of dependence relations. Huge systems are likely to have many conditions and these conditions can be organized into groups. The `get_condition_groups` function returns a list of pairs of the form

   {condition_group_name , [ conditions ] } .

One can use the `condition_group_name` instead of the conditions defined in the list.

We remark that the dependency graph is not necessarily connected. Some modules are not preconditions of any other modules. In this case the definitions of the corresponding conditions are superfluous. Other modules do not have any preconditions, consequently they can be omitted from the return value of the `get_precondition` function.

Let's see an example. Let two applications `app1` and `app2` be given. The first has 3 server nodes that are controlled by a supervisor node. There is another server in the second application which has to wait for the complete startup of the first application. A possible implementation of the above functions might be:

```
...
get_conditions() ->
  [
   { {app1_rootsup   , undefined      } , cond_app1_rootsup },
   { {generic_server , [{app1_server1}]} , cond_app1_server1 },
   { {generic_server , [{app1_server2}]} , cond_app1_server2 },
   { {generic_server , [{app1_server3}]} , cond_app1_server3 }
  ].

get_condition_groups() ->
  [
   { group_app1_app , [ cond_app1_server1,
                        cond_app1_server2,
                        cond_app1_server3,
                        cond_app1_rootsup ] }
  ].

get_preconditions() ->
  [
   { {generic_server  , [{app2_server1}] } , [group_app1_app]  }
  ].
```

### 4.2  The condition server

The startup is controlled by a special server, called `condition_server`, which is started during the Erlang main system start. It stores and handles the dependency graph of the user programs. It also finds and loads the `release_graph` module and checks the validity of the data in it (checks for mistypes, not existing condition names, etc.). Clearly, any error in the `Args` fields remains undiscovered. If the `Args` tags are all `undefined` then the dependency graph is independent from the dynamic data. In this case, an acyclic dependency graph assures dead-lock free structure if each node that has preconditions is started in a concurrent way.

The `condition_server` performs the following two tasks based on the dependency graph. (1) First, sets the conditions belonging to the {M,A}s to true. This is implemented in the `set_condition({M,A})` function. (2) Second, it blocks the caller process until all its preconditions are satisfied. This is imple-mented in the `wait_for_conditions({M,A})` function. These functions have to be called by the generic parts of the behaviours (independently of the users' programs). Consequently, the `condition_server` must be implemented without the `gen_server` behaviour.

We remark that for those modules which don't have any preconditions or don't belong to any other module's precondition, the corresponding function call has no effect.

The `condition_server` module has to be a part of the Erlang kernel modules, since during the Erlang system's startup several event handler and server modules are started, and they require access to the condition storage system.

### 4.3   Modification of the supervisor behaviour

During the startup of a concurrent system, execution fork points must be named. In our case, these places are in the supervisor nodes. We modified the supervisor behaviour so that it accepts extended child specifications. The extension holds an additional field which can be `sequential` or `concurrent`. In the former case, the meaning of the child specification is equivalent with to original one. In the latter case, the supervisor node starts the child concurrently (fork point). We remark that the modification of the supervisor behaviour clearly accepts the original child specifications. The following example shows an extended child specification:

```
{app2_server1, {generic_server, start_link, [app2_server1]},
    permanent, 10, worker, [generic_server], concurrent}.
```

If the generic part of supervisors interprets a concurrent child specification it starts a dummy supervisor node with the proper parameters instead of the original child.

### 4.4   Further modifications of the Erlang system

It is also necessary to modify each Erlang behaviour before the callback `init` function is called, and after it returns successfully. We put these modifications into the `gen_server`, `gen_event`, `gen_fsm`, `supervisor_bridge` and `supervisor` behaviour.

The built-in utilities create boot scripts which do not start the `condition_server` automatically. In order to start the server, a new line has to be inserted into the boot script. The second line of the following code segment shows this:

```
...
  {kernelProcess,heart,{heart,start,[]}},
  {kernelProcess,condition_server,{condition_server,start,[]}},
  {kernelProcess,error_logger,{error_logger,start_link,[]}},
...
```

## 5 Implementation and measurements

We fully implemented the prototype described in the previous sections. The implementation can be used as an OTP extension. This extension is based on Erlang/OTP R11B version and the modifications affected the `stdlib`'s (v. 1.14.1) behaviour modules (namely: `gen_server`, `gen_fsm`, `gen_event`, `supervisor_bridge`, `supervisor`). You can download the prototype from the following url: http://compalg.inf.elte.hu/projects/startup .

Up to now, we described a parallel and reliable solution of the concurrent startup. Our solution gives a well-defined interface for handling the dependency problems among the concurrent starting modules. Therefore it preserves the reliability. It means that reliability of the concurrent startup is based on the dependency graph description of the users' programs. In the following we focus on the running time of the start-up.

We lack access to large industrial applications therefore we created programs for measuring the start-up time in several cases. For simplicity, no dependence conditions were defined, but concurrent supervisor child starting was performed. The measured programs use our modified Erlang/OTP libraries for making fast startup. The tested systems have some `gen_server` and some `supervisor` nodes. The `gen_server` nodes perform time-consuming, resource-intensive computations in their `init` functions. Each measured system has been started both sequentially and concurrently, the time is given in seconds. Each measurement has been performed five times and the figures show the average measured values. The measurements were performed on an SMP 4 machine with 2 AMD Dual Core Opteron, 2GHz, 16 GB RAM, Linux, Erlang 5.5, OTP R11B.

Three different system topologies were measured, a system with (1) deep process tree, (2) wide process tree, and (3) random process tree. The deep process tree was a 3-regular tree of depth 6, the wide process tree was a 10-regular tree of depth 2, and the random process tree was generated using uniform distribution from the range $[1, 5]$ for the number of children of a node, then truncating the tree at level 5.

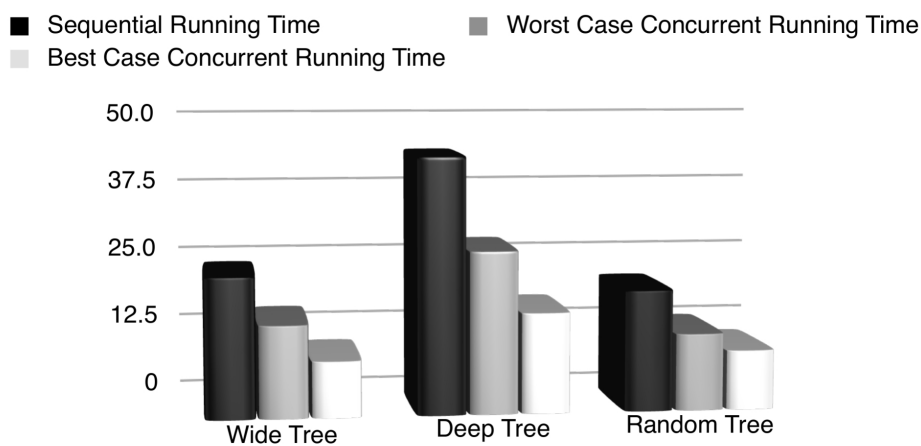

Figure 2: Start-up speed of the sequential and concurrent versions (in seconds). In concurrent case we can put different number of fork points into different places in the process tree. The authors created several concurrent cases for each kind of trees. The worst and best case values represent the slowest and the fastest concurrent start-up time in the proper kind of trees.

We measured the time that is needed for the system start-up as all servers and supervisors were started. The timer started when the `erl` shell was called and stopped when the last server or supervisor started. For this purpose we created a special application which starts just after all other servers or supervisors, and then immediately performs an illegal statement. Since this node crashes at once, consequently `erl` terminates. In other words, we measured the time between the starting and crashing of the Erlang shell.

There are several ways to make a system's process tree concurrent. We tagged the modules which have to start parallel. The speed of the startup depends on the number of the concurrent processes. The deeper the position of the fork point in the tree, the more parallel threads are created (more dummy supervisors). Therefore we show the running times as a function of the number of concurrent threads and as a function of the depth of fork points.

Figure 2 shows that the concurrent versions (not surprisingly) are always faster than the sequential ones. In some cases however, the concurrent start-up was two times faster than the sequential one.

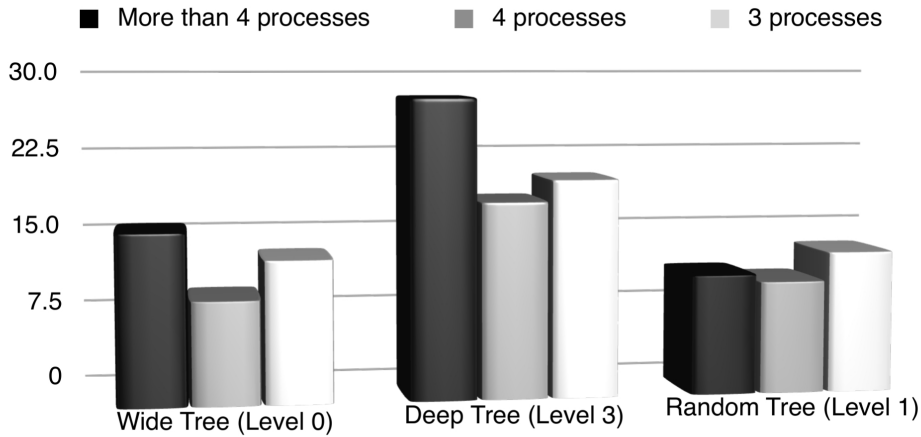

Figure 3: Start-up speed (in seconds) depending on the number of running processes, which can be set by the fork points. The levels show the depth of the fork points.

Figure 3 shows the startup speed as a function of the number of fork points. Since there were 4 processors (2 dual core) in the testbed, it is not surprising that 4-fold parallelism yields the best results. When only 3 parallel processes were started, one processor did not work, and 3 processors performed the whole startup. In case of more than 4 active processes, the processors had to switch between the active processes resulting in a serious overhead. Note however, that the most significant overhead in our measurements comes from the time consuming part of the servers' `init` functions.

Figure 4 shows how the results depend on the depth of the fork points. We measured a fall back performance when all nodes in a given level were started parallel. In this case the system had more concurrent processes in the deeper levels. One can also observe that the version of 4 active process forking is the most resistant to the depth. In this case the only overhead comes from the number of dummy supervisor trees. The measurement suggests that the system should be forked as close to the root as possible.

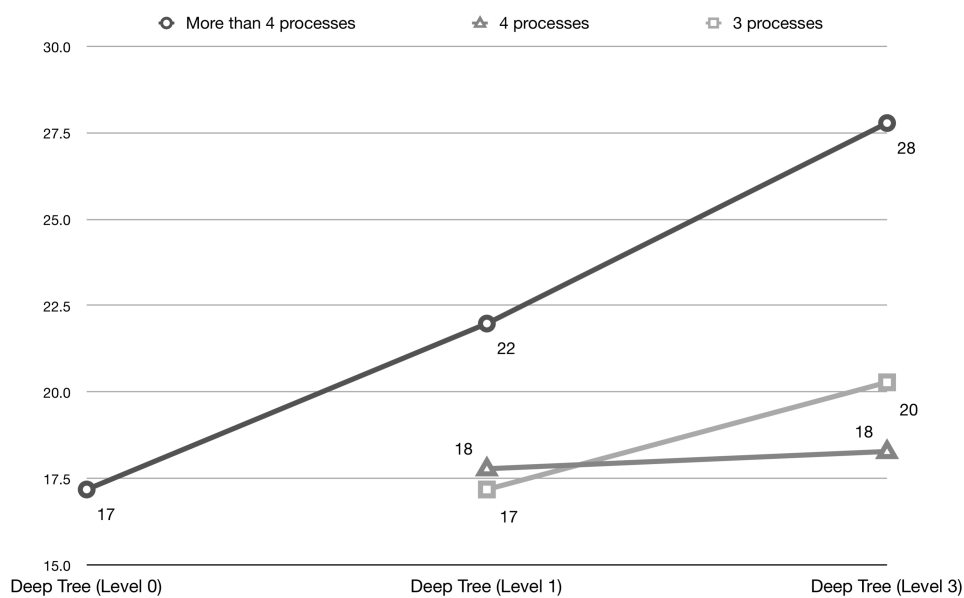

Figure 4: Start-up speed (in seconds) depending on the depth of the fork points.

# 6   Conclusions

In this paper we presented a solution for the parallel start-up of Erlang systems. We gave a general description of the solution and we measured the start-up time in several cases. Our measurements show that the parallel start-up can be much faster than the sequential. On the other hand our solution provides a well-defined mechanism for controlling the dependency relations among processes resulting reliable systems. The main advantages of our solution are:

- Precise and concise dependency handling.
- Preserving the supervision tree structures.
- The dependency graph is an Erlang module.

- The dependency graph is dynamic.
- Less than 150 lines modification in the stdlib.

Disadvantage of our solution is that bad dependency graph could result deadlock or system crash. We conclude that our solution is highly capable for the parallelization of Erlang systems' startup in case of legacy systems and new developments as well.

## 7 Acknowledgements

The authors would thank to the Lemon project (https://lemon.cs.elte.hu) that allowed us to use the project's main computer for measuring. We would also thank to Péter Nagy at Ericsson Telecommunications Hungary for answering our questions about Erlang and OTP.

The research was supported by the project TÁMOP-4.2.1/B-09/1/KMR-2010-003 of Eötvös Loránd University.